\documentclass[%
 aip,
 jmp,%
 amsmath,amssymb,
 reprint,%
]{revtex4-2}

\usepackage[version=4]{mhchem}
\usepackage{graphicx,color,mhchem}
\usepackage{dcolumn}
\usepackage{bm}
\usepackage{placeins}

\begin{document}


\title[black Si]{Heating of black-Si by microwaves at 2.45 GHz: effect of nano-needle orientation}

\author{Massimiliano Zamengo$^{1,*}$,  
Haoran Mu$^{2,3}$, Hsin-Hui Huang$^{2,3}$, 
Tomas Katkus$^{2}$, Darius Gailevi\v{c}ius$^4$, 
Saulius Juodkazis$^{2,4,5}$, 
Junko Morikawa$^{1,5,6,*}$
}

\affiliation{School of Materials and Chemical Technology, 
The Institute of Science Tokyo, 2-12-1, Ookayama, Meguro-ku, Tokyo 152-8550, Japan}
\affiliation{Optical Sciences Centre, School of Science, Swinburne University of Technology, Hawthorn, Victoria 3122, Australia}
\affiliation{Melbourne Center for Nanofabrication (MCN), 151 Wellington Road, Clayton, Vic 3168, Australia}
\affiliation{Laser Research Center, Physics Faculty, Vilnius University, Saul\.{e}tekio Ave. 10, 10223 Vilnius, Lithuania}
\affiliation{~World Research Hub (WRH), School of Materials and Chemical Technology, Institute of Science Tokyo, 2-12-1, Ookayama, Meguro-ku, Tokyo 152-8550, Japan}
\affiliation{~Research Center for Autonomous Systems Materialogy (ASMat), Institute of Innovative Research, Institute of Science Tokyo,Yokohama 226-8501, Japan}
\thanks{Correspondence: zamengo.m.82d4@m.isct.ac.jp (M.Z.);  J.M. morikawa.j.4f50@m.isct.ac.jp }
\date{\today}

\begin{abstract}
Silicon, which is highly reflective and has negligible absorption at the microwave 2.45 GHz range, can be heated when it is turned into black-Si with a surface texture of nano-needles made by \ce{SF6}/\ce{O2} plasma etching. Microwave heating of flat Si and of black-Si with vertical and with tilted (oriented) nano-needles was compared under identical conditions in a cylindrical \ce{TM010} resonator. Samples were placed horizontally on a 150-$\mu$m-thick cover glass, which standardized their height and position inside the cavity and allowed a common calibration of the radiation thermometer against a K-type thermocouple. Under a stepwise protocol of only 2-3-4 W of microwave power (60~s per step), black-Si with vertical needles reached $\sim 260^\circ$C, systematically exceeding flat Si ($\sim 190^\circ$C), while black-Si with oriented needles showed a delayed but abrupt onset of heating at 3 W. In situ monitoring of the cavity resonance revealed a 2-3 MHz downshift of the resonance frequency f for all samples, with the drift consistently appearing once the sample temperature exceeded $\sim 120-150^\circ$C, consistent with thermally activated carrier generation in Si. At a given temperature, each sample loads the cavity at a different f and quality factor Q, indicating an orientation-dependent effective permittivity of the nano-needle array, in line with the strong form birefringence of tilted black-Si. Application potential and challenges in the quantitative temperature and permittivity determination are discussed.
\end{abstract}

\maketitle
\tableofcontents

\section{Introduction}

Black-Si has a black appearance due to its non-reflective ($R<0.5\%$) property defined by the surface texture of nano-needles. Low reflectance $R$ leads to strong absorbance $A\rightharpoondown 1$ when the sample is optically thick and has no transmittance $T\rightarrow 0$. This is valid for the optical near-IR spectral region for which the spacing and height of nano-needles is at sub-wavelength dimensions. For microwave $\sim 2.45$~GHz (12.24~cm wavelength) Si is reflective and cannot be heated due to the negligible loss tangent $\tan\delta = \varepsilon^{"}/\varepsilon^{'}$, where the permittivity $\varepsilon^* = \varepsilon^{'} +i\varepsilon^{"}$; also, it is the dissipation factor $\tan\delta =1/Q$, where $Q$ is the quality factor. Since Joule heating of electrically conductive Si is ineffective, other mechanisms such as interface or surface scattering, can be invoked to increase energy losses, leading to heating. By adding nanoparticles, e.g., SiC, it is possible to heat them by microwave and, consequently, to heat materials in their vicinity~\cite{Massi}. If material is non-absorbing and reflective to $\mu$-waves such as Si, it cannot be heated without adding absorbing SiC nanoparticles.  For electrically conductive materials, metals and semiconductors, the permittivity is described by the Drude model  $\varepsilon^{*}\equiv [n(1-i\kappa)]^2$ and $\varepsilon^{*} = \varepsilon' - i[4\pi\sigma/\omega]$, where $\sigma$ is conductivity, $\omega = 2\pi\nu$ is the cyclic frequency~\cite{SPIE}. Hence, from the real and imaginary parts of the permittivity~\cite{SPIE}: $\varepsilon' = n^2(1-\kappa^2) = n^2\left[1-\left(\frac{\sigma}{n^2\nu}\right)^2\right]$ and $\varepsilon" = 2n^2\kappa = 2\sigma/\nu$. 
 For water at room temperature of 20$^\circ$C at 2.45~GHz, the permittivity is $\varepsilon^{'} +i\varepsilon^{"}\approx 79+i10$~\cite{grapes}. Then, $n = 8.909$ and $\kappa = 0.063$ and $\varepsilon^* = [8.909(1-i0.063)]^2$. Reflectance coefficient at normal incidence (from air) $R = \frac{(n-1)^2 +(n\kappa)^2}{(n+1)^2 +(n\kappa)^2}= 63.8\%$. The conductivity $\sigma = 5\times 2.45\times 10^9$~1/s or 1.36~[S/m] or, in terms of resistivity, $\rho\equiv 1/\sigma = 73.5~\Omega$cm; here SI-to-Gaussian(ESU) unit change 1~S/m~$\equiv 9\times 10^9$~1/s is used. When $\kappa\rightarrow 1$, the $\varepsilon^{'}\rightarrow 0$, i.e., the condition $0< \varepsilon^{'}< 1$ named epsilon-near-zero (ENZ), which corresponds to the efficient energy deposition and absorption. It is noteworthy that when the real part of permittivity decreases to zero $\varepsilon^{'}\equiv 0$, the dielectric breakdown occurs (this is the definition of the breakdown). At such conditions, the 
 microwave frequency $\nu = 2.45$~GHz becomes equal to the conductivity (measured in 1/s in Gaussian units) reduced by a factor $n^2\approx 81$ as calculated according to $\nu = \sigma/n^2$. 
 
The microwave heating is an established technology~\cite{Osep} which has become a household utility for heating food based on the vibration and collision of the lightweight \ce{H2O} dipoles. Industrial scale heating systems~\cite{polym} for material processing are well established~\cite{Horikoshi2024}. High temperature synthesis of SiC from graphite and Si was proposed by microwave heating~\cite{poland}. Catalysts for oxygen reduction reaction (ORR) based on  Pt-M alloys (where metal M = Fe, Co, Ni) can be made using microwave heating~\cite{zhao}. Another green energy application for superior quality production of \ce{LiFePO4} powders for batteries is by use of processing  based on microwave heating~\cite{LiB}. Carbon materials are among best microwave absorbing materials and are used as heaters in increasing range of industrial material processing substituting conventional oven heating~\cite{carb}. 

Microwave annealing has gained renewed attention as a low-thermal-budget alternative to conventional rapid thermal processing, with recent applications spanning Si photovoltaics, ferroelectric memory, and plasmonic nanostructure fabrication. In Si solar cells, $\mu$-wave annealing achieves effective hydrogen passivation of polycrystalline Si passivating contacts in only 1-2 minutes, matching the performance of 30~min conventional nitrogen annealing~\cite{Truong2025}. For ferroelectric devices, $\mu$-wave annealing of Hf$_{0.5}$Zr$_{0.5}$O$_2$ capacitors at $500\,^{\circ}$C yields a remanent polarization of $63\,\mu$C\,cm$^{-2}$ substantially higher than the $40\,\mu$C\,cm$^{-2}$ obtained by rapid thermal annealing at the same wafer temperature along with suppressed wake-up effects and reduced leakage current~\cite{Zhao2022}. Microwave heating also drives controlled morphological evolution in metallic thin films: Au films on BK7 dewet into well-defined nanoislands above $550\,^{\circ}$C with faster and more uniform heating than oven annealing~\cite{Bercea2025}, and microwave absorption in ultra-thin Au films can promote out-of-plane $\langle111\rangle$ crystallographic alignment regardless of substrate~\cite{Kim2025}. The technique has further been applied to crystallise (Na$_{0.5}$Bi$_{0.5}$)TiO$_3$ thin films for tunable microwave dielectric devices~\cite{Joseph2025}. Together these results highlight microwave annealing as a rapid, selective, and low-thermal-budget processing route applicable to a broad class of functional materials.

Here, we compare heating of Si by 2.45 GHz microwaves when its surface is textured with nano-needles, i.e., turned into black-Si by dry plasma etching 
Three samples were investigated under identical conditions: flat Si, black-Si with vertical nano-needles (sub-500 nm separation and height) and black-Si with tilted (oriented) nano-needles obtained by angled plasma etching 
All samples were measured in the same horizontal geometry, resting on a thin cover glass which fixed their height inside the cavity and made the calibration of the radiation thermometer common to all experiments. It was found that black-Si is heated markedly faster and to higher temperatures than flat Si at microwave powers as low as 2-4 W, and that the needle morphology affects the heating transients as well as the evolution of the cavity resonance frequency f and quality factor Q. When replotted against the sample temperature, the drift of f and Q reveals a thermally activated change of the dielectric response above $\sim 120-150^\circ$C which is common to all samples, while the sample-specific values of f and Q at a given temperature point to an orientation-dependent effective permittivity of the nano-needle array.

\begin{figure*}[b!]
\centering\includegraphics[width=1\textwidth]{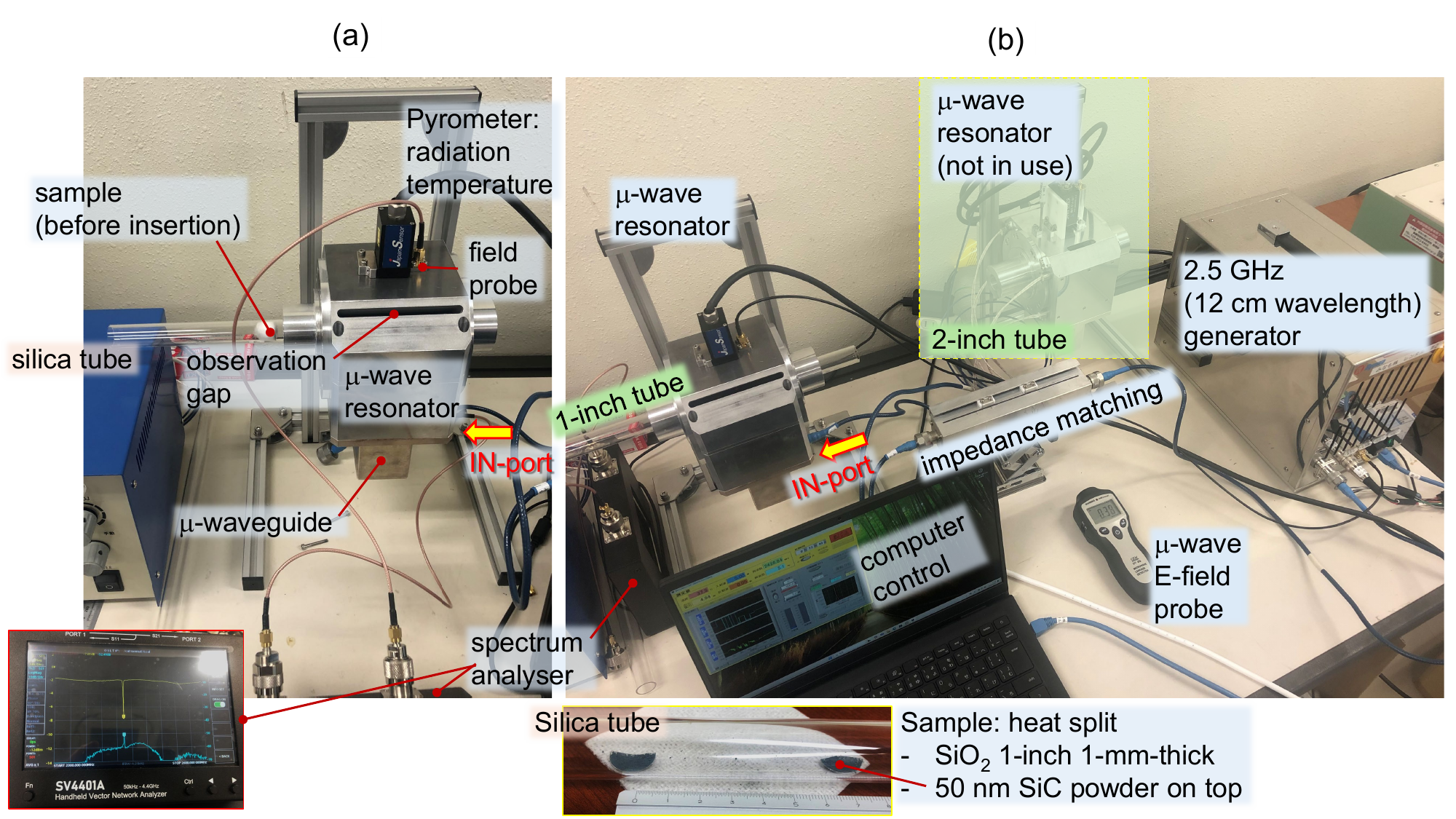}
\caption{\label{f-fig1_rev} Setup of microwave heater based on generator, microwave resonator/cavity and pyroelectric radiation thermometer. (a) Setup connected in a spectral analyser mode, which visualise cavity mode and changes to it (bottom-inset). 
(b) Setup in heater operation mode; outside microwave leaking detector was always on.}
\end{figure*}

\section{Experimental: samples and methods}

\subsection{Black Si}

Black-Si (b-Si) is produced by dry plasma etching using a simple \ce{SF6}/\ce{O2} gas mixture~\cite{16aplp076104,20n873,16semsc221}. 
The main results are shown for Boron-doped p-type Si was used as substrate for plasma etching of b-Si. The energy level of B is 45~meV above the valence band $E_v$ of Si; 45~meV energy is equivalent to 250$^\circ$C. Black-Si with vertical nano-needles was obtained by a 40~min etch, which results in needles of $\sim 400$ nm height and sub-500 nm separation, perpendicular to the Si surface ($\theta = 90^\circ$). Black-Si with tilted (oriented) nano-needles was made by plasma etching with the Si substrate placed at an angle during the etch; the resulting tilted nano-needles ($\theta = 60^\circ$ from the surface to the needle axis) define a large $\Delta n = 0.45$ form birefringence at the visible spectral range~\cite{20oe16012}. The following three samples were compared (Figure 2): reference flat Si ($8.7 \times 8.6 \times 0.5$ mm$^3$), b-Si with vertical needles ($7.6 \times 8.1 \times 0.3$ mm$^3$) and b-Si with oriented needles ($6.9 \times 5.8 \times 0.5$ mm$^3$); the flat-Si reference was cut to a size comparable to that of the b-Si samples.
It is informative to compare the critical plasma density $n_{cr} = \omega^2\varepsilon_0m_e^{*}/e^2$ with Si doping levels, here $\varepsilon_0$ is the permittivity of free space, $m_e^{*} = 1\times m_e$ is the effective mass of electron, which was taken equal to the actual electron mass, $e$ is the charge of electron, $\omega=2\pi\nu$ is the cyclic frequency. For $\nu = 2.45$~GHz, one finds $n_{cr} = 7.446\times 10^{10}$~cm$^{-3}$, i.e., even undoped Si has electron density $n_e\gg n_{cr}$ and strong reflection is expected. 

\subsection{Microwave heating at 2.45~GHz}

Microwave heating and dielectric property measurements were carried out using a cylindrical TM010 mode resonator operating from 2.4 GHz to 2.5 GHz (Minamo Co. Ltd., Japan). The resonator was connected via a waveguide, a slug tuner, and coaxial cables to a microwave signal generator (MR-2G-100, Ryowa Electronics Co. Ltd., Japan), with impedance matching and frequency tracking managed by dedicated software to maintain resonance. The resonant frequency $f$ [MHz] and quality factor $Q$ were obtained by fitting the VNA data with Lorentzian and Gaussian functions (pseudo-Voigt fitting). Temperature was monitored with a radiation thermometer (TMHX-STM0050, Japan Sensor Co. Ltd., Japan) calibrated against K-type thermocouples (Fig. A1).

Figure~\ref{f-fig1_rev} shows the 
used setup~\cite{Massi} capable of launching up to 100~W power into the central cavity at 2.45~GHz (12.24~cm wavelength, 0.01~meV, 0.0817~cm$^{-1}$ wavenumber). Temperature was monitored directly on the 
Si sample (through the \ce{SiO2} tube). 
Radiation temperature was determined from 1.95-to-2.6~$\mu$m wavelength range using InSb sensor; temperature can be determined up to $1000^\circ$C. The sensing area diameter on the sample was 3~mm at the 7~cm depth into the cavity from the rim (5~mm at 5~cm and 8~mm at 10~cm depths).   
The \ce{SiO2} tube was open to room air access during all experiments (without forced flow).
The \ce{SiO2} tube had an inner diameter of 16 mm. Samples of different lateral size and thickness, when laid directly on the curved bottom of the tube, sit at slightly different heights, intercept different local intensities of the E-field of the \ce{TM010} mode, and present the pyrometer measurement spot at slightly different positions between runs. For a fair comparison between materials, the sample height inside the cavity was therefore standardized: all samples were placed horizontally on a 150-$\mu$m-thick cover glass laid at the bottom of the SiO$_2$ tube (Figure 2). The cover glass did not measurably change the cavity resonance conditions. In this way, samples of different sizes are exposed to a consistent E-field and the pyrometer sensing spot falls at the same position for every sample, which allows a single temperature calibration to be applied to all experiments (Fig. A1). For the b-Si sample with oriented needles, the in-plane projection of the needle axis was set either parallel or perpendicular to the E-field by rotating the sample about the tube axis by 90$^\circ$.

The heating protocol consisted of three consecutive 60 s steps of constant microwave power: 2 W, 3 W and 4 W (Figure 2). Such Watt-level powers were sufficient because of the improved and reproducible coupling of the horizontally mounted samples with the cavity mode. The microwave setup was controlled by a computer program. The heating of the sample was in situ monitored by a sensor for the constantly changing permittivity in order to fine tune the frequency for the best energy coupling into the cavity. The frequency tunability window was $\pm 50$ MHz around the center frequency of 2.45 GHz.

A vector network analyzer (VNA; model SV4401A, Sysjoint Co. Ltd., China) is used to monitor resonance conditions upon insertion of the sample into the cavity (Figure 1(a)). When the sample has a broad resonance that partially overlaps with the cavity resonance, the sample can be heated up. Samples that reflect incoming energy from the waveguide back into the generator cannot be heated up and have to be reduced in size to mitigate large reflectance.



\begin{figure*}[b!]
\centering\includegraphics[width=0.95\textwidth]{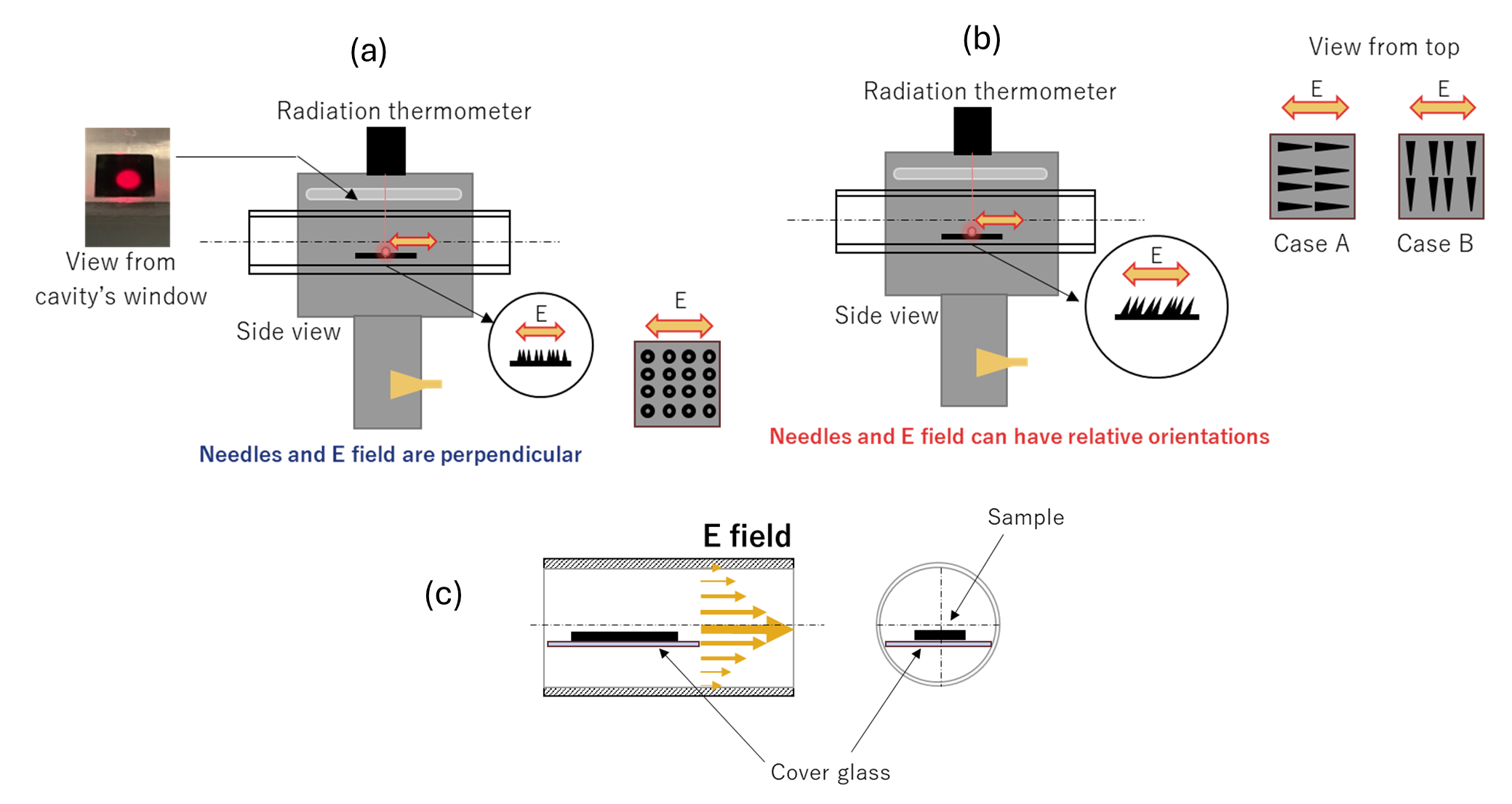}
\caption{\label{f-bSiN}  Schematic of the sample mounting geometry and needle-to-electric field orientation inside the microwave cavity. (a) Sample with vertical b-Si needles mounted horizontally inside the quartz tube; the electric field $E$ is perpendicular to the needles. The diagram shows the pyrometer alignment spot viewed through the cavity window. (b) Sample with oriented b-Si needles; by rotating the sample, the in-plane projection of the needle axis is set parallel (Case A) or perpendicular (Case B) to $E$. (c) Cross-sectional view of the quartz tube showing the sample resting on a 150-$\mu$m -thick cover glass to ensure consistent positioning at the maximum electric field strength.
}
 
\end{figure*}
\begin{figure*}[tb]
\centering\includegraphics[width=.95\textwidth]{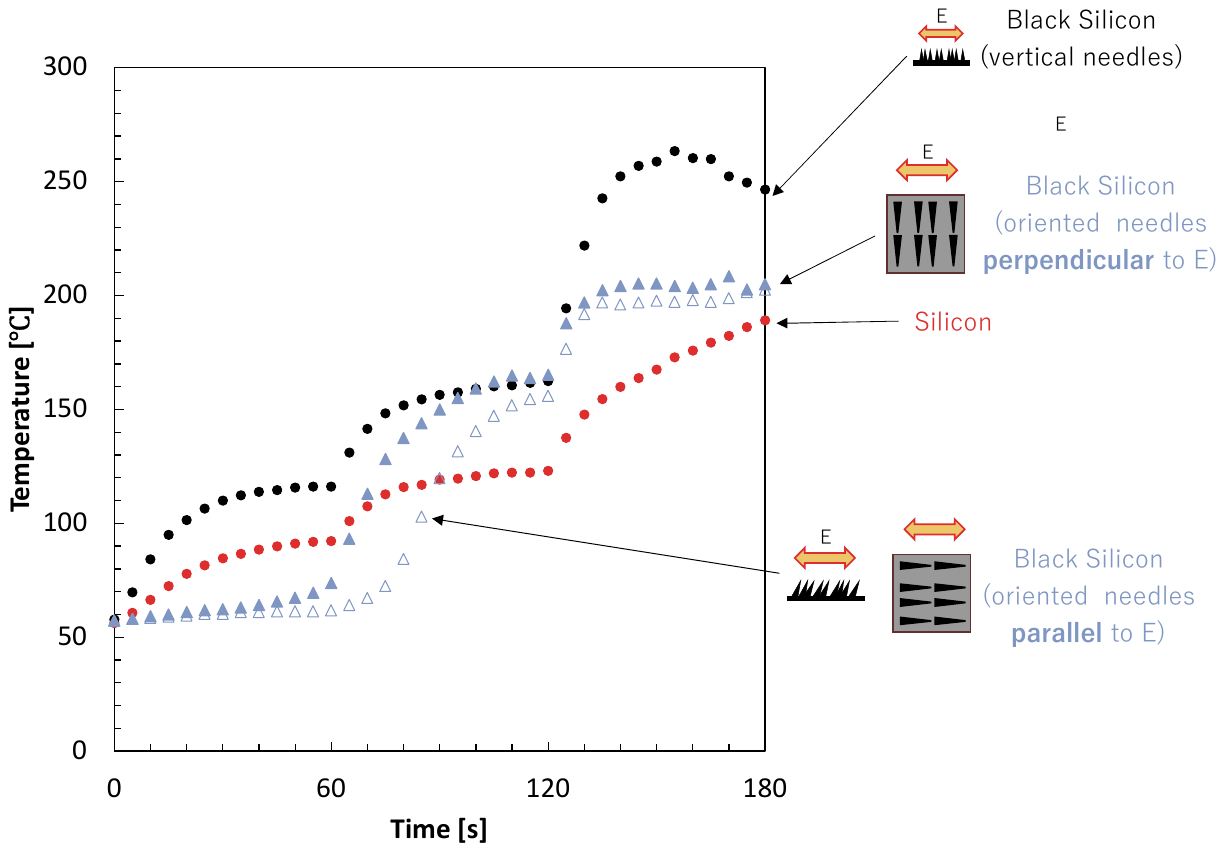}
\caption{\label{f-res} Time evolution of the surface temperature during the 2-3-4 W protocol (60 s per step) for flat Si, b-Si with vertical needles and b-Si with oriented needles (in-plane needle projection parallel and perpendicular to $E$). Temperature was measured by the calibrated radiation thermometer. b-Si with vertical needles reaches $\approx 260,^\circ$C at 4 W; the oriented-needle sample heats slowly at 2 W and rises abruptly upon switching to 3 W, overtaking flat Si.
}
\end{figure*}

\section{Results and discussion}

\subsection{Heating of flat Si and black-Si }

Figure 3 shows the temperature evolution measured on the surface of the three samples during the 2-3-4 W protocol. Black-Si with vertical needles was heated fastest and to the highest temperature, reaching $\approx 260,^\circ\mathrm{C}$ at the 4 W step. The reference flat Si heated more slowly under the same program, reaching $\approx 190,^\circ\mathrm{C}$. The b-Si sample with oriented needles heated slowly at 2 W but showed a sudden temperature rise upon switching to 3 W, overtaking flat Si and reaching $\approx 205,^\circ\mathrm{C}$ at 4 W; the largest temperature jumps of all samples were found upon the switch to 4 W. Within the reproducibility of the experiment, the heating transients of the oriented-needle sample were essentially the same for the in-plane needle projection set parallel or perpendicular to the $E$-field. The systematically faster and stronger heating of both b-Si samples demonstrates the additional loss channel introduced by the nano-needle texture: free charges driven by the microwave $E$-field inside needles of nanoscale cross-section experience increased interface/surface scattering, which enhances the Joule losses and heating.

\subsection{Frequency and quality factor evolution under microwave exposure }

During heating, the resonance frequency generally dropped by $\sim 2$--$3$ MHz for all the samples (Figure 4(a)). The onset and steepness of the drift, however, differed between samples: for b-Si with vertical needles and for flat Si the drift developed gradually during the 3 W and 4 W steps, while for the oriented-needle sample $f$ remained nearly constant at 2 W and then dropped abruptly at the 3 W step, concomitantly with the sudden temperature rise seen in Figure 3. The $Q$-factor evolution was also sample specific (Figure 4(b)). Under the 4 W step, $Q$ increased for b-Si with vertical needles while it decreased for flat Si. For b-Si with oriented needles, a sudden drop of $Q$ was observed at 3 W, after which $Q$ recovered and became larger than its initial value. These trends show that the different black-Si morphologies couple differently with the cavity as their temperature changes. The overlay of the cavity spectra acquired during the experiments (Figure 5) visualizes these trends: flat Si shows a progressive sharpening and a downward frequency shift of the resonance; b-Si with vertical needles shows peak broadening together with the downward shift; the oriented-needle sample starts with sharp, tall peaks whose amplitude suddenly collapses at the onset of fast heating and then grows again. No significant difference was observed in the spectral evolution when the in-plane orientation of the oriented needles was changed with respect to the $E$-field direction.

\begin{figure*}[tb]
\centering\includegraphics[width=.99\textwidth]{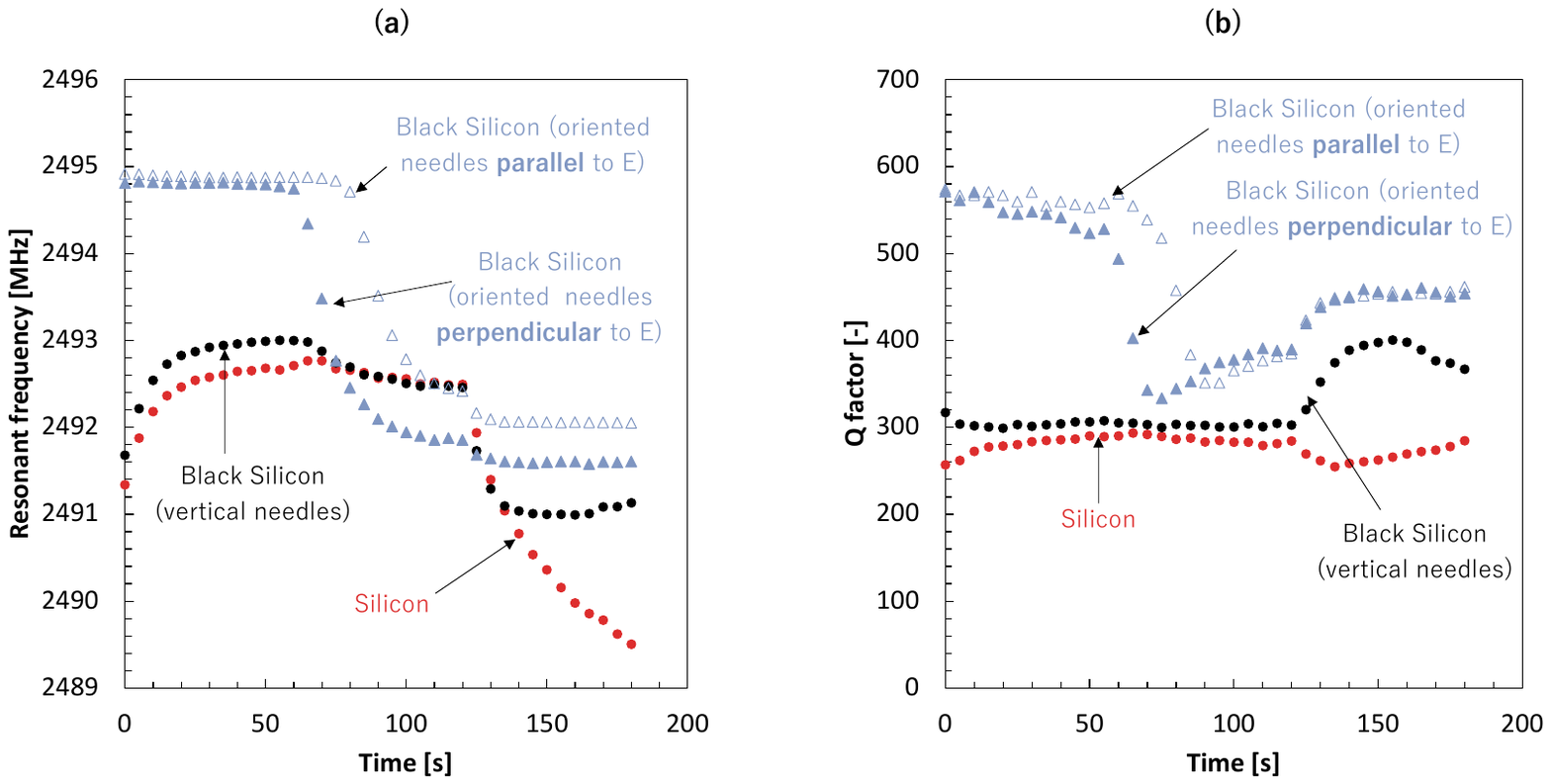}
\caption{\label{f-fig4_rev} (a) Resonance frequency $f$ and (b) quality factor $Q$ of the cavity versus time during the heating protocol for the four configurations of Figure 3. $f$ drops by $\sim 2$--$3$ MHz for all samples; under the 4 W step, $Q$ increases for b-Si with vertical needles while it decreases for flat Si; the oriented-needle sample shows a sudden collapse of $Q$ at the 3 W step followed by a recovery.
 }
\end{figure*}
\begin{figure*}[tb]
\centering\includegraphics[width=1\textwidth]{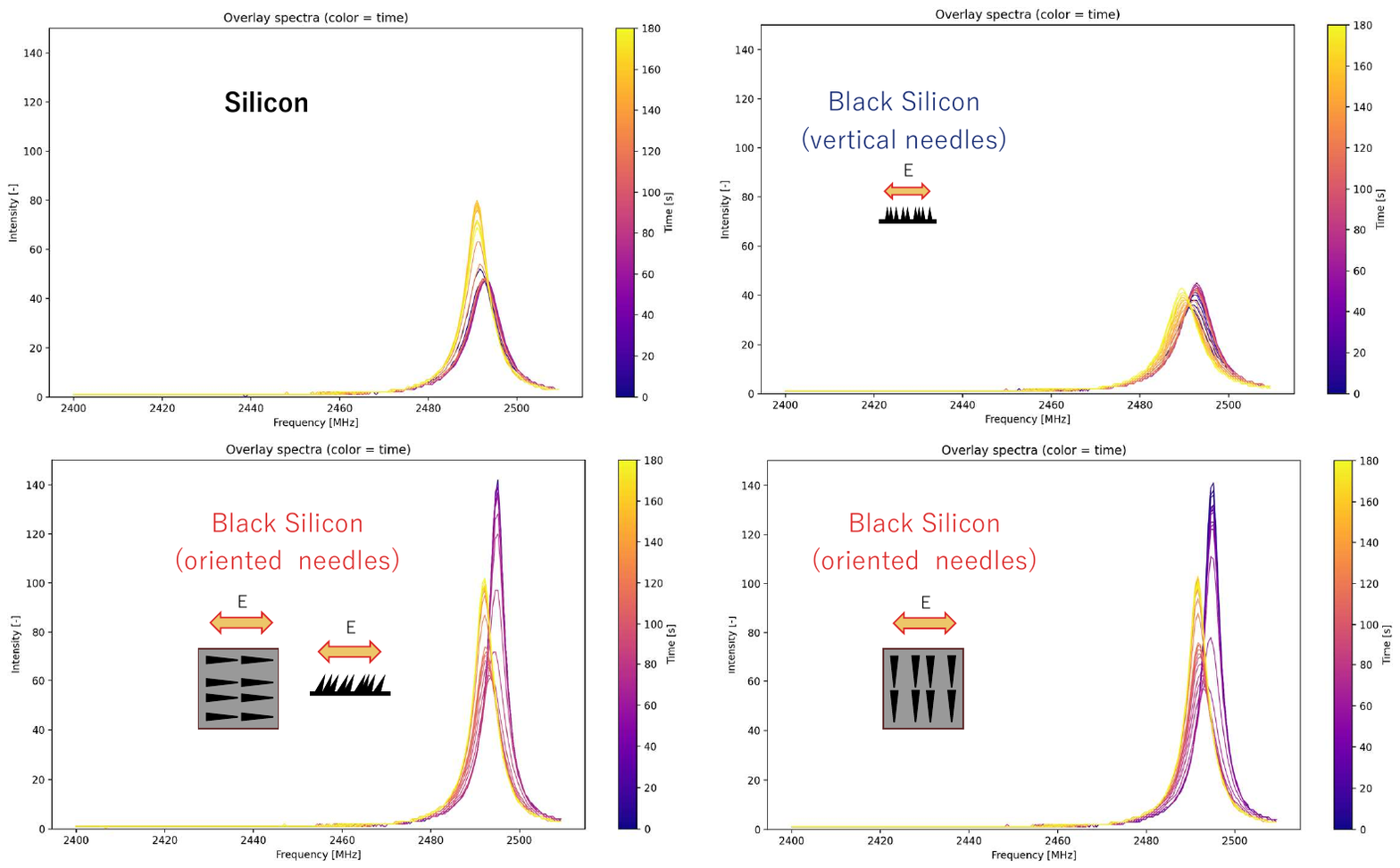}
\caption{\label{f-fig5_rev} Overlay of the cavity spectra (color encodes time, 0-180 s) for flat Si, b-Si with vertical needles and b-Si with oriented needles in the two in-plane orientations. Flat Si shows progressive sharpening and a downward frequency shift; b-Si with vertical needles shows peak broadening and a downward shift; the oriented-needle sample starts with sharp peaks whose amplitude collapses at the onset of fast heating and then grows again.}
\end{figure*}

\subsection{Temperature dependence: thermally activated carriers}\label{disco}

Figure 6 gives the overall summary of the experimental results replotted as temperature dependences of the resonance frequency and the $Q$-factor. Although the $f$-drift occurs at different times during the power protocol for each sample (Figure 4), when the data are replotted versus the sample temperature the drift consistently appears once the sample reaches approximately 120--150 $^\circ$C, indicating a temperature-activated change in the dielectric properties which is common to all samples. The decrease in resonance frequency indicates that the effective dielectric loading of the cavity increases during heating. Since the resonance frequency of a TM$_{010}$ cavity is highly sensitive to the real part of the permittivity, the observed shift toward lower frequencies suggests an increase in the effective dielectric polarization of the samples at elevated temperature. Strong temperature dependence of $f$ at $(150 \pm 50),^\circ$C is consistent with the $E_a = 45$ meV thermal ionization of the B-acceptors responsible for the free-carrier (holes) increase of conductivity $\sigma = ne\mu$, where $n$ is the carrier concentration, $e$ is the elementary charge, and $\mu$ is the mobility. The carrier density increases as $n \propto e^{-E_a/(2kT)}$, where $k$ is the Boltzmann constant and $T$ is the absolute temperature. In the temperature window where all B-dopants are ionized, the mobility reduction via phonon/lattice scattering governs the conductivity, $\sigma \propto \mu \propto T^{-3/2}$; at even higher temperatures ($T > 250,^\circ$C) the intrinsic, over-the-bandgap ($E_g = 1.12$ eV) thermal carrier generation starts, causing growth of $\sigma \propto n \propto e^{-E_g/(2kT)}$. The interplay of these mechanisms can produce a non-monotonic evolution of the cavity frequency: the frequency of the cavity mode depends on the effective refractive index $n_0$ of the material inside the cavity of length $L$ as $f = c/(2n_0L)$ (the fundamental mode), where $c$ is the speed of light in free space; a reduction of $n_0$ causes an increase of $f$, and vice versa. The slight initial increase of $f$ up to $\sim 100$--$120^\circ$C followed by the steep decrease above $\sim 150^\circ$C observed for flat Si and for the vertical-needle b-Si (Figure 6(a)) is consistent with such competing contributions of carrier generation and mobility to $\sigma(T)$ and, through it, to the effective permittivity. Importantly, this behavior cannot be explained by a simple emissivity artefact of the pyrometric temperature measurement. Independent calibration experiments performed under conventional (uniform, PID-controlled) heating of the resonant cavity up to approximately 250$^\circ$C did not reveal significant nonlinearities in the temperature calibration curves of any of the three samples (Fig. A1), and confirmed that flat Si and b-Si have the same emissivity within the measurement accuracy. Although microwave irradiation could still modify the apparent emissivity through non-equilibrium thermal or electronic effects, emissivity variations alone cannot account for the simultaneous and systematic evolution of both the resonance frequency and the cavity $Q$-factor. Moreover, the measured $f$, $Q$ dependences versus temperature were reversible and repeatable; hence, there was no oxidation of Si, which would alter the emissivity permanently.

\begin{figure*}[tb]
\centering\includegraphics[width=1\textwidth]{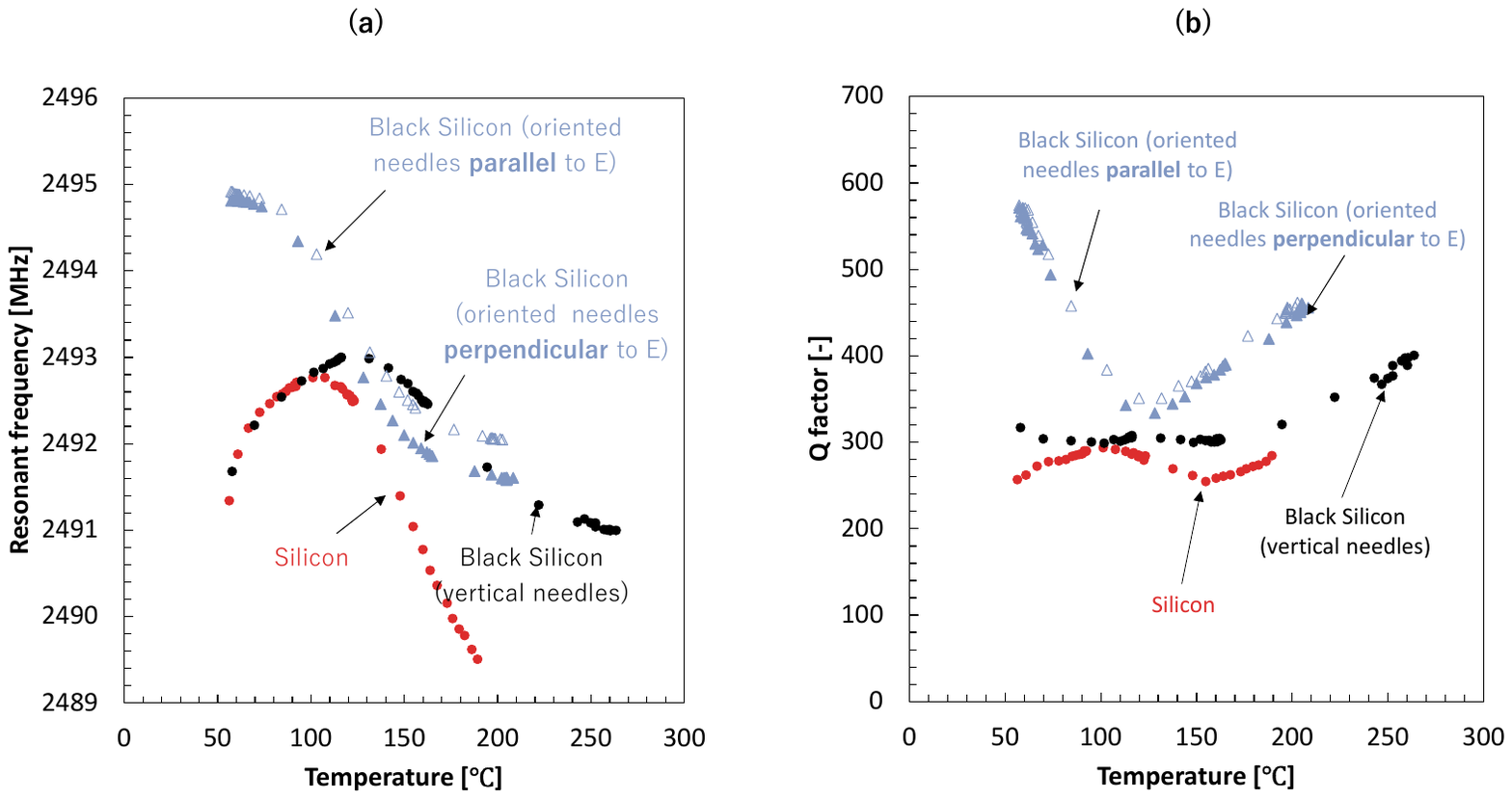}
\caption{\label{f-fig6_rev} (a) Resonance frequency $f$ and (b) quality factor $Q$ replotted versus the sample temperature for the four configurations. The drift of $f$ consistently sets in at $\sim 120$--$150,^\circ$C for all samples; at a given temperature, each sample loads the cavity at a distinctly different $f$, revealing the sample- and orientation-specific effective permittivity.
}
\end{figure*}

\subsection{Orientation-dependent effective permittivity of the nano-needle array }

A remarkable observation in Figure 6(a) is that, at a given temperature, each sample loads the cavity at a distinctly different resonance frequency: the oriented-needle b-Si perturbs the cavity least (highest $f$), while flat Si and the vertical-needle b-Si produce a stronger dielectric loading. Since all samples were mounted at the same height and measured with the same calibration, this sample-specific offset reflects genuine differences in the effective permittivity of the samples and, in particular, in the orientation of the nano-needle layer relative to the $E$-field. The nano-needle layer can be treated as a uniaxial effective medium with the optic axis along the needle direction. For an array of high-aspect-ratio needles of volume fraction $f_v$ in air, the depolarization factor along the needle axis is $L_{\parallel} \approx 0$, while across the needles it is $L_{\perp} = 1/2$ (infinite-cylinder limit). The two principal values of the effective permittivity then follow the Wiener bound for $E$ parallel to the needles, $\varepsilon_{\parallel} = f_v\varepsilon_{\mathrm{Si}} + (1 - f_v)$, and the Maxwell Garnett expression for $E$ perpendicular to them, $\varepsilon_{\perp} = [(1 + f_v)\varepsilon_{\mathrm{Si}} + (1 - f_v)]/[(1 - f_v)\varepsilon_{\mathrm{Si}} + (1 + f_v)]$. Taking $\varepsilon'_{\mathrm{Si}} \approx 11.7$ and $f_v \approx 0.3$ gives $\varepsilon_{\parallel} \approx 4.2$ and $\varepsilon_{\perp} \approx 1.7$, i.e., the needle layer is strongly anisotropic, consistent with the giant form birefringence $\Delta n = 0.45$ measured for tilted black-Si in the visible spectral range.~\cite{20oe16012}

Because the same mixing rules apply to the imaginary part of the permittivity, the dissipated power density $w = \frac{1}{2}\omega\varepsilon_0\varepsilon''_{\mathrm{eff}}|E|^2$ inherits this anisotropy. For the vertical-needle sample in the horizontal mounting, the $E$-field of the \ce{TM010} mode lies in the sample plane, perpendicular to the needle axis, and only $\varepsilon_{\perp}$ is probed; for the tilted-needle sample, $E$ acquires a finite projection onto the needle axis which depends on the in-plane orientation of the sample. The fact that the two in-plane orientations gave very similar heating and resonance evolution suggests that, at 2.45 GHz (a wavelength five orders of magnitude larger than the needle spacing) the response is dominated by the out-of-plane component of the needle orientation distribution and by the thermally activated carrier dynamics, rather than by the in-plane anisotropy alone. Orientation-dependent (angular) electromagnetic responses have been reported for vertically aligned carbon-nanotube arrays in the microwave-to-THz range~\cite{cite16},  
and anisotropic effective-medium descriptions are well established for Si nanowire arrays~\cite{cite17};  
the present results extend these concepts to the thermal (heating) response of nanotextured Si at 2.45~GHz.

A plausible explanation of the $Q$-factor evolution is related to the anisotropic electromagnetic response of the b-Si needle array. When the electric field is oriented perpendicular to the needles, strong localized electric fields and depolarization effects are generated in the narrow regions between adjacent structures. At low temperature, these highly localized fields may produce significant scattering, field singularities, and localized dissipative losses, resulting in a comparatively low cavity $Q$-factor. Indeed, for this orientation, the $E$-field is concentrated in the air-gaps since the normal component of the displacement $\varepsilon'E$ is continuous across the Si--air interface, as follows from the boundary conditions; the tangential component of the $E$-field is continuous across the boundary. As the temperature increases, thermally activated carriers within the silicon may partially screen these localized depolarization fields. This screening effect can reduce local field confinement and suppress hotspot-related dissipation, leading to a more spatially homogeneous electromagnetic response inside the cavity. Consequently, microwave losses decrease and the cavity $Q$-factor increases, as observed for the vertical-needle b-Si above $\sim 150$-$200,^\circ$C (Figure 6(b)). Simultaneously, the increased carrier density and polarization capability of the heated silicon structure may enhance the effective real permittivity of the material without necessarily increasing dissipative losses. In this regime, the b-Si behaves less like a strongly scattering micro-structured surface and more like an effective anisotropic dielectric medium with enhanced polarizability. 

This interpretation naturally explains the simultaneous decrease in resonance frequency and increase in $Q$-factor observed for the vertical-needle sample. The abrupt collapse of $Q$ for the oriented-needle sample at the 3 W step, followed by its recovery, is consistent with the sample being swept through the condition of maximum loss tangent (analogous to the epsilon-near-zero condition discussed in the Introduction) as its conductivity rises with temperature: dissipation first increases up to the matching point and then decreases as the sample becomes more screening/reflective. The relatively large cavity perturbation produced by a relatively small sample further supports the hypothesis of exceptionally strong electromagnetic coupling between the microwave field and the b-Si morphology. Overall, the results suggest that b-Si under microwave irradiation behaves as a temperature-dependent anisotropic electromagnetic medium whose dielectric polarization and dissipative response evolve dynamically with both temperature and field orientation. These observations may provide insight into the microwave behavior of structured semiconductors and metamaterial-like silicon surfaces, particularly in regimes where morphology-induced field localization and thermally activated carrier dynamics become strongly coupled, e.g., laser-patterned  \ce{VO2}~\cite{23apl093101} where the metal-to-dielectric transition occurs at $\sim 70^\circ$C.

The effective permittivity extracted from the standard resonance description can only serve as a semi-quantitative estimate, since the exact size (volume) of the sample, its placement relative to the on-axis maximum of the $E$-field of the cavity mode, and the thermal contact with the cover glass and the SiO$_2$ tube all affect the instantaneous thermal conditions. However, a qualitative understanding of the sample heating is achieved, and it shows the possibility of an in situ feedback on material response/changes in real time via observation of $f$ and $Q$. Importantly, heat treatment and annealing protocols can be applied with higher precision and without latency, e.g., for colloidal particle precipitation and growth or structural defect annealing in glasses and crystals.

\section{Conclusions and Outlook}



Nanotexturing of the Si surface with nano-needles strongly enhances the coupling of 2.45 GHz microwave energy into the sample. With all samples mounted horizontally at a standardized height on a thin cover glass, black-Si with vertical needles was heated to $\approx 260,^\circ$C by only 4 W of microwave power, systematically above the flat-Si reference, and black-Si with tilted (oriented) needles showed an abrupt onset of heating at 3 W. The evolution of the cavity resonance frequency $f$ and quality factor $Q$, monitored in situ, traces the changes of the sample permittivity in real time: all samples show the onset of the $f$-drift once their temperature exceeds $\sim 120$--$150,^\circ$C, consistent with thermally activated carriers, while the sample-specific $f$ and $Q$ values at a given temperature reveal the anisotropic, orientation-dependent effective permittivity of the nano-needle layer. Nanotextured surfaces could be promising for microwave heating applications instead of nanoparticles, e.g., SiC, which is more difficult to control from being airborne. High thermal conductivity of Si is useful to heat samples/materials which do not absorb microwaves when they are placed in close contact. Anisotropy in heat deposition is a new and largely unexplored direction promising for externally controlled response of micro-patterns and structures useful for solely EM-controlled micro-devices.


\small\begin{acknowledgments}
M. Z. acknowledges support via MEXT Next-Generation Computational Science Grand Reach Program (Grant No.JPMXP1920260105). S. J. acknowledges support via the Australian Research Council  DP240103231 grant. 
J. M. acknowledges support via JST CREST (Grant No. JPMJCR19I3) and KAKENHI (No.26K01228).
\end{acknowledgments}

\bibliography{aipsamp}
\appendix
\setcounter{figure}{0}\setcounter{equation}{0}
\setcounter{section}{0}\setcounter{equation}{0}
\makeatletter 
\renewcommand{\thefigure}{A\arabic{figure}}
\renewcommand{\theequation}{A\arabic{equation}}
\renewcommand{\thesection}{A\arabic{section}}

\section{Temperature calibration}

\begin{figure*}[h!]
\centering\includegraphics[width=.55\textwidth]{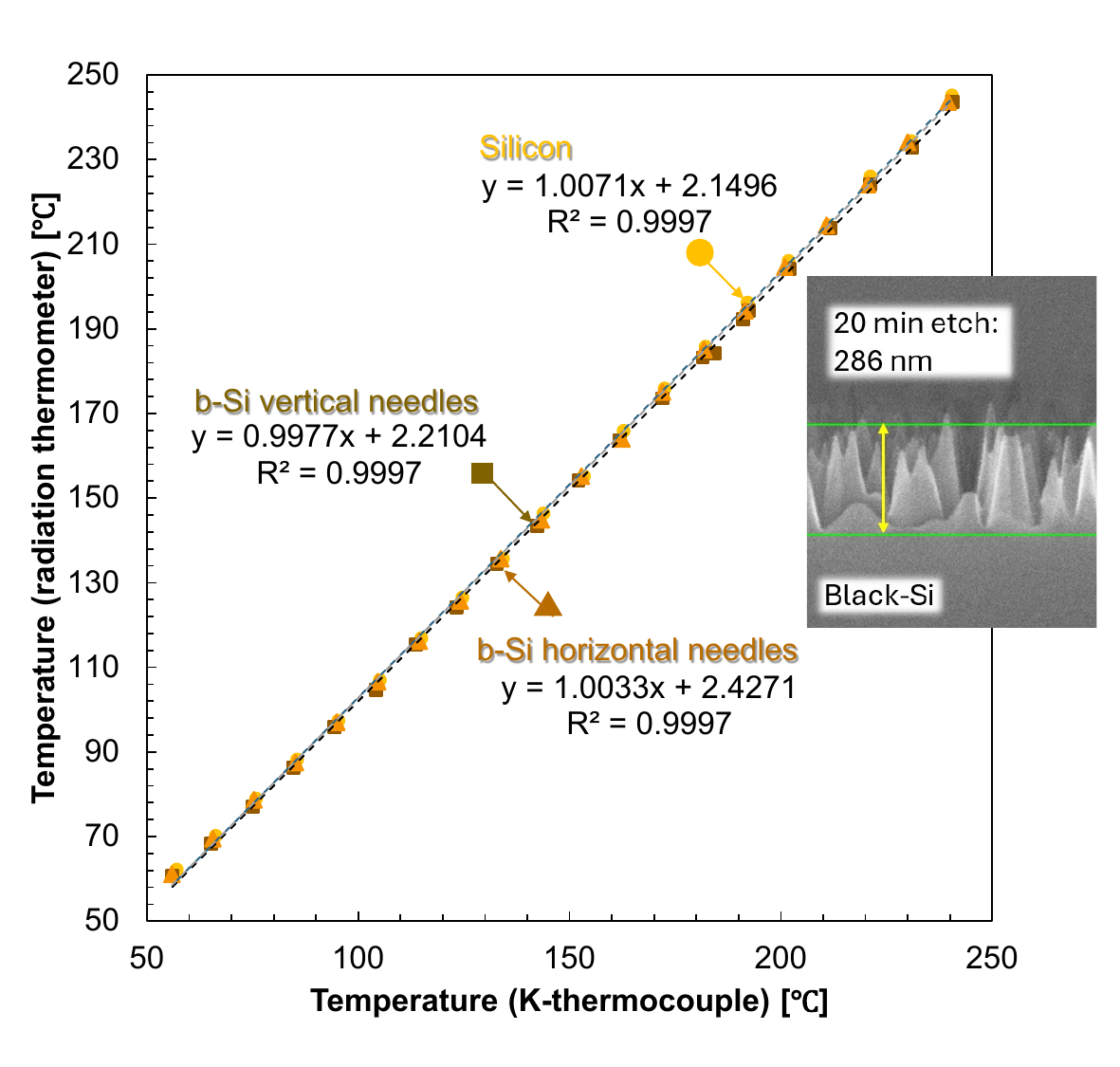}
\caption{\label{f-tk} Calibration of the radiative pyrometer using a standard type-K
Chromel-Alumel thermocouple. Black-Si (b-Si) is plasma etched sample with nanoscale needle-like pyramidal structures~\cite{13oe6901,16semsc221}. The inset shows scanning electron microscopy (SEM) image of the side-view cross section of the black-Si sample with $\sim 290$~nm tall needles; orientation in the image corresponds to the ``vertical needles'' in the heating experiment.}
\end{figure*}

\begin{figure*}[h!]
\centering\includegraphics[width=1\textwidth]{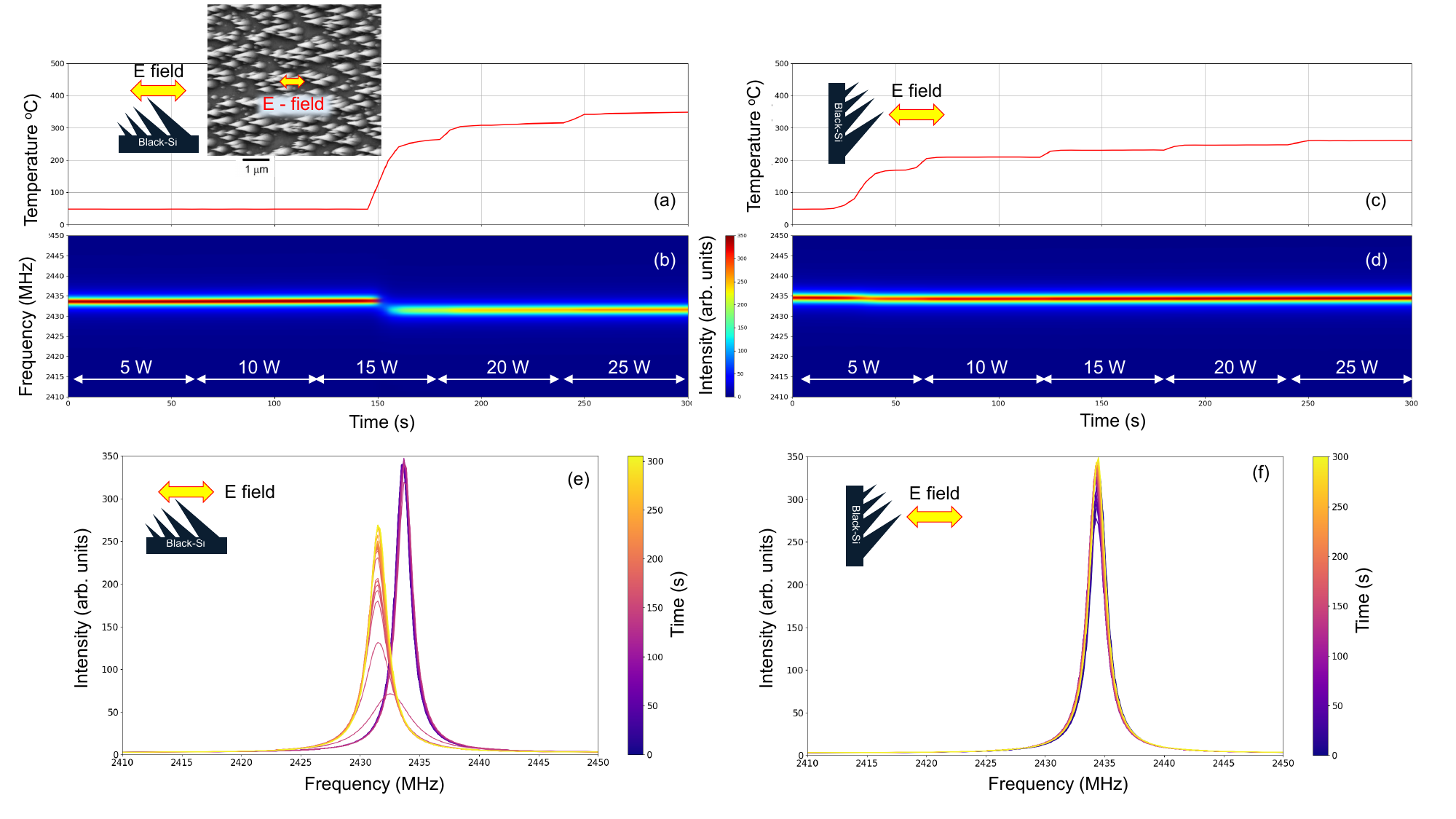}
\caption{\label{f-bSiT}  Time evolution of temperature on tilted black-Si sample placed horizontally and vertically in microwave cavity upon five steps of 5~W power increments for each for 6~s; tilting angle $\theta = 60^\circ$ from surface to needle axis. (a) Temperature vs time was measured by using a radiative pyrometer on the sample surface. Insets show SEM image of tilted black-Si sample in horizontal orientation. (b) Resonance frequency of the cavity during the entire span of the heating experiment; horizontal sample placement (tilted black-Si needles). (c) Temperature vs time for vertical placement of sample (sample was supported by \ce{SiO2} wool). It was noted a high leakage from the cavity triggering alarm of the meter $> 5$~mW/cm$^2$. (d) Resonance evolution during power increase. Sample size was very similar to the one with vertical needles, only it was 0.5-mm-thick. 
Evolution of the resonance for horizontal (e) and vertical (f) placement of tilted black-Si sample.
 }
\end{figure*}
\begin{figure*}[h!]
\centering\includegraphics[width=.99\textwidth]{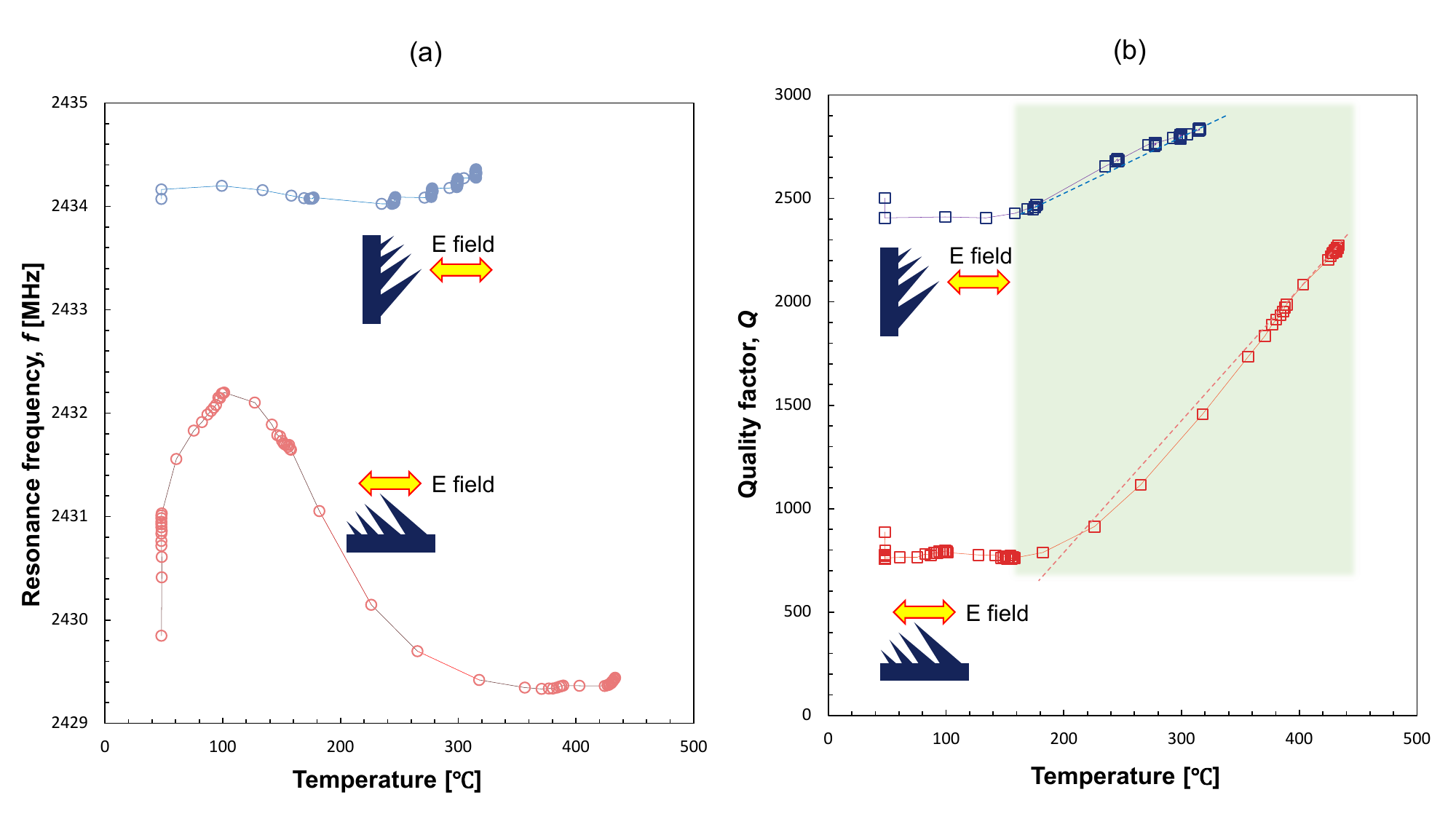}
\caption{\label{f-epsT} Temporal profiles up microwave heating: resonant frequency (a) and quality factor $Q$ 
for horizontally and vertically placed b-Si with tilted needles.  
 }
\end{figure*}

Figure~\ref{f-tk} shows calibration results when b-Si in two orientations as well as flat-Si (not etched but cut to the same size as b-Si). The microwave cavity was heated completely and uniformly using four PID-controlled cartridge heaters. The temperature setpoint was increased in 10$^\circ$C steps at 15-minute intervals, after steady-state conditions had been achieved according to the readings of the type-K thermocouple attached to the sample and the pyrometer. The readout of pyrometer temperature is plotted against temperature by type-K thermocouple. Linearity and high fidelity of the fit $R^2 >0.99$ was confirmed. The measurements were carried out till the 250$^\circ$C. 
This calibration experiment also confirmed that emissivity of nano-smooth surface of Si as well as the nano-structured surface of b-Si are the same.

\section{Heating black-Si with tilted needles}\label{suppl}

A sample of  black-Si with tilted needles was made by plasma etching with Si substrate placed at an angle during plasma etch~\cite{20oe16012}. This resulted at the tilted nano-needles which defined a large $\Delta n = 0.45$ form birefringence at visible spectral range. The sample of very similar size as that with vertical nano-needles was cleaved from the tilted black-Si. The same sample heating and mounting protocol was used as for the black-Si with vertical nano-needles.      

The orientation of the sample strongly affects the resonant condition of the cavity, shifting to lower frequency when sample is positioned horizontally (Fig.~\ref{f-bSiT}). A large change of resonant frequency was observed while heating the sample positioned horizontally. The shift and broadening of the resonant peak suggests a change of properties with temperature. The vertical position of sample causes microwave leakage from the cavity. Figure~\ref{f-epsT} shows temperature dependence of resonant frequency and Q-factor of the b-Si with tilted needles. The remeasured dependencies are very similar to those of b-Si with needles normal to the surface. 

\end{document}